\documentclass[preprint]{emulateapj}

\slugcomment{Accepted for Publication in the Astrophysical Journal Letters}
\shortauthors{Zaritsky \& Psaltis}
\shorttitle{Outer Disks and MOND}

\begin{document}
\title{Outer Galactic Disks and a Quantitative Test of Gravity at Low Accelerations}

\author{Dennis Zaritsky and Dimitrios Psaltis}
\affil{Steward Observatory, University of Arizona, 933 North Cherry Avenue, Tucson, AZ 85721}

\email{dzaritsky@as.arizona.edu, psaltis@as.arizona.edu}
                                                                                                                                       
\begin{abstract}
We use the recent measurement of the velocity dispersion of star-forming, outer-disk knots by Herbert-Fort et al. in 
the nearly face-on galaxy NGC 628, in combination with other data from the literature, to execute a straightforward test of gravity at low accelerations. 
Specifically, the rotation curve at large radius sets the degree of 
non-standard acceleration and then
the predicted scaleheight of
the knots at that radius provides the test of the scenario. For our demonstration,  we presume
that the H$\alpha$ knots, which are young (age $<$ 10 Myr), are distributed like the gas from which they have recently formed and find a marginal ($>$ 97\% confidence) discrepancy with a modified gravity scenario given the current data. More interestingly, we demonstrate that 
there is no
inherent limitation that prevents such a test from reaching possible discrimination at the $> 4\sigma$ level
with a reasonable investment of observational resources.
\end{abstract}

\keywords{gravitation ---  galaxies: kinematics and dynamics}

\section{Introduction}

The application of the Newtonian law of 
gravity to the dynamics of galaxies and galaxy clusters
leads to the well-known discrepancy between the inferred dynamical
mass and the directly observed mass \cite[cf.][]{zwicky,smith,rogstad,rubin}. This discrepancy is, in turn, the motivation for 
the dark matter hypothesis --- that the vast majority of matter in the Universe
is of some undetermined composition. Although the need for some dark matter in certain
environments is now well
established \citep{clowe,angus},
and experiments appear tantalizingly close to directly detecting it \citep{dama}, the possibility that
gravity deviates from Newtonian behavior in the low acceleration regime is not precluded by the
existence of dark matter.

Much of the work on proposed alternatives to Newtonian gravity focus on
demonstrating the consistency of such models with observations, particularly with the measured kinematics of disk galaxies at large galactocentric radii \cite[cf.][]{begeman, mcgaugh}.
The ultimate aim however is to discriminate between the modified gravity 
and dark matter hypotheses, and this requires identifying a regime in which 
the two hypotheses make distinctly different predictions. There are some basic requirements
for whatever environment one
envisions constructing such a test in. First, one will need to work in
a regime well beyond the scale where Newtonian gravity first appears to fail.
Second, one will need to 
work in a regime appropriate for the proposed modified gravity model.
Because the currently proposed modifications to Newtonian gravity are not yet complete theories, 
there is little to be gained by invalidating a specific gravitational variant beyond its 
intended working parameters. Third, one will need to identify an environment in 
which the physical distributions of the baryons and 
hypothesized dark matter are sufficiently different that they create distinct
gravitational signatures. Lastly, one will need to identify a dynamically simple environment
where the interpretation does not depend on one's understanding of a complex dynamical history. 

All of these requirements drive us to study the outskirts of disk galaxies. The observed outer-disk rotation curves provide
one of the principal arguments for deviations from Newtonian gravity \citep[if one assumes no
dark matter, e.g., modified Newtonian dynamics or MOND; ][]{milgrom, begeman, mcgaugh} and
so satisfy the first and second criteria.
The third and fourth criteria are also satisfied in
that the baryons in at least some outer disks have been 
shown to lie in a dynamical cold, undisturbed disk \citep{christlein,herbert-fort}, and the
dark matter, while not in a perfectly spherical distribution, is thought to lie a significantly rounder one \citep{hoekstra, mandelbaum}.

For motions in the disk plane, the modified gravity and dark matter hypotheses are degenerate, with both
predicting larger rotational velocities than inferred from the luminous matter using Newtonian 
gravity. The kinematics in the vertical direction, where the
baryons are strongly concentrated toward the disk plane, but the posited dark matter is not, provides the discrimination between the two hypotheses. 
Such a test was implemented by \cite{sanchez} using the thickness of the H I  
disk at large radii in our own Galaxy.
While they find a modest discrepancy with
the MONDian prediction, they note that the comparison
is somewhat compromised by other potential sources of support for the gaseous layer, such as 
magnetic fields, cosmic rays, and SNe injected energy. Because one needs to find
thicker than predicted H I  to invalidate MOND, and because uncertain sources of 
support will act to thicken the gas distribution, this test has an unresolved weakness. 
We propose to work in the same environment, which has so many appealing features, 
but to utilize another tracer population
that cannot be supported by magnetic fields, cosmic rays, or injected energy. In this {\sl Letter}
we explore the implications of the measured dynamics of outer disk star-forming knots \citep{ferguson, thilker, zaritsky} on modified gravity models.

\section{Testing the MOND Paradigm}

The vertical structure of galactic disks in modified Newtonian
dynamics has been explored by a number of authors (e.g., Famaey \&
Binney 2005; Nipoti et al.\ 2007; S{\'a}nchez-Salcedo, Saha, \&
Narayan 2008).  Using the phenomenological equation (Bekenstein \&
Milgrom 1984)
\begin{equation}
\nabla\cdot\left[\mu\left(\frac{\vert\nabla \Phi\vert}{a_0}\right)
\nabla\Phi\right]=4\pi G\rho
\label{eq:MOND}
\end{equation}
for the gravitational field $\Phi$, where $\mu \left(\frac{\vert\nabla \Phi \vert}{a_0}\right)$ is the 
phenomenological factor that depends on the magnitude of the local acceleration, 
 $g\equiv \vert \nabla \Phi\vert$, in units of the MOND
scale $a_0\simeq 10^{-8}$~cm~s$^{-2}$,  S{\'a}nchez-Salcedo et al.\ (2008)
showed that the disk scale-height in the deep MOND regime is given implicity by
\begin{equation}
{1\over 2}L(x) \left ({v_c\over R} \right)^2 h^2 + \pi {G\over \mu (x)}\Sigma_{\rm L} h - \sigma^2 = 0
\label{eq:MOND_h}
\end{equation}
In the above expressions, $G$ is the gravitational constant,
$\sigma$ is the velocity dispersion of matter in the vertical (out-of-disk plane) direction, $\Sigma_{\rm L}$
is the local column density of luminous (i.e., baryonic) material,  $v_c$ is the circular
velocity of matter, $R$ is the galactocentric radius, 
and $L(x)$ is the logarithmic derivative of $\mu$ at $x$, where $x \equiv g/a_0$. Various fitting functions for $\mu(x)$, with the appropriate limiting behavior, have
been developed, including what is often called the standard one, $\mu = x/\sqrt{1+x^2}$ \citep[cf. ][]{begeman}, and $\mu = x/(1+x)$ \citep{fb}. 

We estimate the magnitude of the phenomenological factor $\mu$ at
the radius of the knots using the equation for the
radial acceleration
\begin{equation}
\mu\left (x \right)v_{\rm c}^2=\frac{G M_{\rm L}}{R}\;,
\label{eq:MOND_radial}
\end{equation}
where $v_{\rm c}$ is the circular velocity of matter, and $M_{\rm L}$
is the enclosed mass of luminous matter. This equation is valid for a
spherical distribution of mass $M_L$. In fact, we are discussing a thin disk of baryonic material
with a surface brightness (and corresponding mass column density) that declines exponentially with 
radius, as is typical for spiral galaxies. For such a mass distribution the
predicted circular velocity is somewhat larger than that of a spherical distribution of equivalent enclosed mass \citep[cf.][]{binney}, which implies an even lower value of $\mu$ and hence a stronger MOND effect. However, for 
large $R$ the difference is small, $<$ 10\% at $R > $  5 disk radial scalengths. 
We confirm, using the standard fitting formula for $\mu(x)$ that equation (3) produces a somewhat
larger value of $\mu$ than the fitting formula (0.38 vs. 0.32, respectively) and therefore a correspondingly somewhat larger value of $h$ --- which is conservative given the argument we are constructing. Combining
equation~(\ref{eq:MOND_h}) and (\ref{eq:MOND_radial}), we obtain
\begin{equation}
{1\over 2}L(x) \left ({v_c\over R} \right)^2 h^2 + {\pi \Sigma_{\rm L} R v_c^2 \over M_{\rm L}} h - \sigma^2 = 0,
\label{eq:test}
\end{equation}
which we solve for $h$. 

The test is then a simple comparison of the predicted and observed values of $h$,
which requires knowledge of the six measurable quantities: $h, M_L, \sigma, R, \Sigma_L, $ and $v_c^2$. 
The more difficult of these to measure are 
$h$ and $\sigma$, and in particular
it is difficult to measure both
quantities for any one galaxy because $h$ is best measured in inclined systems while
$\sigma$ is best measured in face-on systems. One can combine characteristic values 
of $h$ and/or $\sigma$ to stand in place of the value for an individual galaxy in order to 
pursue this test. Of the two quantities, the one that
has been measured in a number of systems is the scale-height, $h$ \citep{obrien}, although only
for the H I layer and not yet for the knots themselves. This shortcoming
must eventually be remedied because otherwise our test is also susceptible to the
unknown heating sources mentioned previously. Because
the test has a stronger sensitivity to $\sigma$, we present a measure
of $\sigma$ in the nearly face-on galaxy NGC 628 and use a representative estimate of $h$ for disk galaxies.

\section{The Data}

We now describe how we estimate each of the quantities necessary to evaluate equation~(\ref{eq:test}).

\subsection{The Disk Scaleheight, $h$}

The only currently available measurement of the scaleheight of outer disks comes from 
measurements of the H I flaring \citep{sancisi, merrifield, olling, obrien}. These of course (in the current
context) have the same weakness described by \cite{sanchez} in that the thickness may be
due to effects other than thermal pressure. Nevertheless, the currently definitive study of \cite{obrien}
shows that the typical scaleheight at the radii of interest here is $\sim $ 1 kpc. 
Their results also show a universal, steep rise in $h$ toward the edge of the H I disk, suggesting that this
test can become increasingly powerful at larger radii, or that additional heating is occurring near the edge of the neutral gas disk. Of the galaxies studied, IC 2531 is the nearest
analog to NGC 628. They are both classified as Sc in NASA/IPAC Extragalactic Database. The V band magnitudes are $-$20.9 and $-$20.2, respectively. The H I extents
are in both cases $\sim$ 30 kpc and the rotation speeds are 220 and 170 km sec$^{-1}$. 
At the corresponding mean radius of our
observations (18 kpc; see below), $h$ = 1.2 kpc. The value of $h$ 
as a function of radius for the two galaxies in their sample with circular velocity $v_c > 100$ km s$^{-1}$ is the same, except
for the final steep rise just before the end of the H I data. The uncertainty, judging from the 
scatter in $h$ as a function of $R$ and a comparison of the two more massive galaxies in 
the sample, appears quite small ($ \sim 10$\%). The more serious, but not easily quantified,
uncertainties relate to how well these galaxies describe NGC 628 and how well does the
H I scaleheight describe that of the knots. With a typical velocity of 11 km sec$^{-1}$,
H$\alpha$-emitting knots will
travel about 100 pc in their lifetime, significantly less than one scale height. 

The future of this test lies in directly measuring the scaleheight of the star forming knots. 
This can be done with either deep GALEX or H$\alpha$ observations of edge-on galaxies, but
has not yet been attempted. The value of the GALEX knots is that in their lifetime ($>$ several
hundred Myr) they will have 
completed several vertical oscillations and will therefore test whether the H I and knots settle
into the same vertical distribution.

\subsection{The Enclosed Mass of Luminous Matter, $M_L$}

The luminosity of the galaxy ($M_V = -20.73$, based
on $m_v = 9.25$ (NED) and our adopted distance of 9.9 Mpc) is $1.6 \times 10^{10} L_\odot$ for
$M_{V,\odot} = 4.79$. For $B-V = 0.51$ and from the relationships for stellar mass-to-light 
ratios from \cite{bell}, we estimate $M/L_V  = 1.9$ and calculate that the stellar mass is $3.1 \times 10^{10} M_\odot$. The uncertainty in this value will be dominated by that in $M/L$. The integrated magnitude 
is not likely to be off by more than 0.1 mag, and the scatter in distance models (NED) corresponds to 0.1 mag as well. As a result the uncertainty in the luminosity cannot exceed 20\%. 
However, $M/L$ could in principle be off by substantially more. \cite{bell} estimate, both from the
differences obtained with different models and the scatter in their Tully-Fisher relationship, that 
the random scatter in $\log M/L$ is 0.2 (or about 50\% in $M/L$, which is the uncertainty we adopt for $M_L$). To this mass we add the gas as measured by \cite{kamphuis}. They measured 
a total H I mass of $12 \times 10^{10} M_\odot$, of which they attribute $3 \times 10^{10} M_\odot$ to 
the distorted disk that lies beyond our measured knots. When we then correct the remainder 
for the presence of He by multiplying by 1.4, we estimate a gas mass of $12.6 \times 10^{10} M_\odot$. 
The total mass we use is then $43.6 \times 10^{10} M_\odot$, with an uncertainty of 50\%. 

\subsection{The Velocity Dispersion, $\sigma$}

The observational data consist of the radial velocity measurements of star-forming
regions in the outer disk of the nearly face-on galaxy NGC 628. Star forming regions generate
strong emission lines (H$\alpha$ was used in this case) whose redshifts can be measured precisely. 
We currently have a measurement by \cite{herbert-fort} of the vertical velocity dispersion in the outer
disk of NGC 628 using 14 such regions at galactocentric radii ranging from
1.04 to 1.79 $R_{25}$ (13 to 23 kpc).  There are three principal sources of uncertainty
in this measurement. First, the sample size is small. This could,  in principle, be 
corrected and does not pose an inherent limit on the usefulness of this test. We have an estimate
of this random uncertainty, $\sim 4 $ km sec$^{-1}$, from the work of \cite{herbert-fort}. Second, there is an uncertainty
regarding the orientation and kinematic parameters of the disk about which we are
calculating the dispersion. Again, \cite{herbert-fort} attempted to explore this uncertainty 
and concluded that for several of the more plausible orientations the uncertainty in $\sigma$ was about
2 km s$^{-1}$.  For the current purpose we adopt an uncertainty of 4 km s$^{-1}$, but 
caution that there could be a systematic problem. This possibility should be addressed both with more
data on this galaxy and with observations of other galaxies. 

A more fundamental limitation of the current data is that we measure the velocity dispersion
only for knots that are relatively young ($<$ 10 Myr) because we require the presence of the 
H$\alpha$ emission line. This results in
our using knots that are closely related to the neutral hydrogen, and which have had insufficient time to erase the possible
influence of unknown heating sources on the gas. The eventual solution to this problem is to measure the scaleheight of both this young population of knots and of the older GALEX population. If these two populations share a similar scaleheight, then the concerns of using a young tracer population are alleviated. Otherwise, 
we will need to somehow measure the velocity dispersion of the older, fainter knots.

\subsection{The Radius, $R$}

Due to the sparseness of the velocity field sampling, we must use data spanning a range
of radii. The mean radius of the \cite{herbert-fort} data is 18 kpc. The uncertainty in this comes
from the uncertainty in the distance, which we estimate to be 10\%.  The more significant, and
unquantifiable uncertainty comes from applying this test to data over a range of radii. In
the future, a denser sampling of the kinematics will allow us to mitigate against potential systematic
errors arising from this averaging (as well as enabling a stronger test by examining the radial
behavior of any discrepancies in equation~[\ref{eq:test}]). 

\subsection{The Column Density of Luminous Matter, $\Sigma_L$}

Over this same region there are independent 
measurements of both the gaseous and stellar mass surface densities for NGC 628\footnote{The numerical values presented here differ from those in \cite{herbert-fort} because we have `conservatively' chosen to adopt a longer distance, which mitigates the discrepancy between the MONDian prediction and observations. }.
The H I gas,  measured via the 21 cm line emission \citep{kamphuis}, has a mass of $\sim 6\times 10^9 M_\odot$
over the relevant radial range, resulting in an estimated
mean surface gas mass density of 5.6 $M_\odot\ {\rm pc}^{-2}$ after we multiply by 1.4 to account for He. This estimate of the gas mass 
excludes molecular gas, which we know is 
present at some level since there is ongoing star formation. 
The stellar surface density can be estimated from the measured
surface brightness at $1.3 R_{25}$ \citep[$\mu_V \sim 27$ mag arcsec$^{-2}$;][]{natali}
and an estimated $M/L_V$ of $3$ for an old stellar population \citep[for $B-V = 1$ and a modified 
Salpeter IMF, see][]{bell} to be 1.0 $M_\odot\ {\rm  pc}^{-2}$.
The surface mass density, $\Sigma_L$, is therefore likely to be $> 6.6 \ M_\odot\ {\rm pc}^{-2}$ at the representative radius of $1.3 R_{25}$. 

The uncertainty in this value is difficult to quantify in part because the data are not ours
and in part because it is likely to be dominated by systematic errors. Because the
H I mass dominates the sum, the principal source of uncertainty is likely to come from that measurement. We use the mean surface density evaluated 
over a large area. It would clearly be preferable to have more local measurements of both
$\sigma$ and $\Sigma_L$ to avoid such averaging. We will assign the rather
large uncertainty of 30\% in an attempt to capture these possible errors. Ultimately, this
test will require a more refined measurement of $\Sigma_L$ and its uncertainty.

\subsection{The Rotational Velocity, $v_c$}

The rotation curve is poorly measured because of the galaxy's low inclination. 
The rotational velocity at the outermost measured radius, $\sim$ 12 kpc, is
$\sim 180$ km s$^{-1}$ with no evidence for a decline with radius \citep{fathi}, although the velocity derived from their fitted model at the same radius
is 150 km s$^{-1}$. 
The H I study by \cite{kamphuis} adopted
a flat rotation curve with amplitude 200 km s$^{-1}$ and found good agreement with their data (except
for a couple of high velocity clouds). We adopt $v_c$ = 170 km s$^{-1}$ with an uncertainty of 
20\%.

\section{Comparison to Predictions}
  
Solving equation~(\ref{eq:test}) using the discussed measured values
for NGC~628, and accounting for uncertainties using Monte-Carlo techniques, we find that $h$ is predicted to be $356^{+316}_{-235}$ and $298^{+274}_{-172}$ kpc for the standard and Famaey-Binney interpolation formulae, respectively. The observed value of 1.2 kpc is rejected with
slightly $>$ 97\% confidence in both cases. In a scenario with Newtonian gravity and in which 
the dark matter is 
not concentrated in the disk, so that the dominant mass component is the observed baryons,  the
predicted scaleheight, given by $h = \sigma^2 /G \pi \Sigma_{\rm L}$, is $1280^{+560}_{-426}$ kpc,
which is in agreement with the observations.

There are two ways in which we can improve the fidelity of this test.
First, we can concentrate on lowering the uncertainties. Several of the terms contribute
comparably to the final error, although the error on $\sigma$ is currently the largest. It would
be possible to reduce this uncertainty by observing more H$\alpha$ knots and using somewhat higher
resolution observations to minimize the importance of the instrumental contribution to the
observed deviations. For comparison, the H I velocity dispersions for the outer disks of the
\cite{obrien} sample
are all $\sim$ 7 km s$^{-1}$. These results suggest, but do not demonstrate, that
the knot $\sigma$'s might be somewhat lower than our measurement.  The next largest contribution
comes from the uncertainty in the rotation velocity. An analysis of the H I kinematics in which
the rotation is solved for would help address this issue.  Second, we can focus at larger
radii, where the radial behavior differences of the modified gravity and dark matter scenarios would provide additional discriminatory power. 
The difficulty will be measuring $\sigma$ well in a small
annulus since the H$\alpha$ knots are somewhat rare. A realistic expectation is that we could
reduce all of the uncertainties, except in the distance, by a factor of two given a feasible investment of current-day observational resources. In that case, the discrepancy (for the current parameter values) would exceed the 99.99\% confidence level.

\section{Conclusions}

We present the first use of the kinematics of outer disk star forming knots as a test of 
modifications to gravity at low accelerations.
We evaluate the prediction for the disk scaleheight in modified Newtonian gravity \citep{sanchez} and compare to the typical H I scaleheight of the outer disk in similar galaxies. For our estimated uncertainties, we find 
a marginal ($>$ 97\% confidence) discrepancy between modified gravity scenarios and the
observed scaleheight and excellent agreement between a model with Newtonian gravity and
a baryonic disk. There remain various aspects of this test 
that require additional observations to confirm. 
In particular measurements of the vertical scaleheight of the H$\alpha$-emitting knots and of
the older GALEX-identified knots, would directly address several potential 
systematic uncertainties in 
the current version of the test. If those questions are satisfactorily resolved, we conclude that with
additional data, this test has the potential to provide strong constraints ($> 4\sigma$) 
on existing alternative gravity models with a relatively modest investment of observational resources.

\acknowledgments{DZ acknowledges financial support from NSF (AST-0907771) and 
NASA (LTSA NNG05GE82G) and DP from the NSF CAREER grant (NSF 0746549). This research has made use of the NASA/IPAC Extragalactic Database (NED) which is operated by the Jet Propulsion Laboratory, California Institute of Technology, under contract with the National Aeronautics and Space Administration.}


\begin{thebibliography}{}

\bibitem[Angus et al.(2007)]{angus}
Angus, G.W., Shan, H.., Zhao, H.S., and Famey, B. 2007, \apjl, 654, 13

\bibitem[Bell \& de Jong(2001)]{bell}
Bell, E.F. and de Jong, R.S. 2001, \apj, 550, 212

\bibitem[Begeman, Broeils, \& Sanders(1991)]{begeman}
Begeman, K.G., Broeils, A.H., \& Sanders, R.H. 1991, \mnras, 249, 523

\bibitem[Bekenstein \& Milgrom(1984)]{1984ApJ...286....7B} Bekenstein,
  J., \& Milgrom, M.\ 1984, \apj, 286, 7

\bibitem[Bernabei et al.(2008)]{dama}
Bernabei, R., et al. 2008, Eur. Phys. J. C 56, 333

\bibitem[Binney \& Tremaine(1987)]{binney}
Binney, J. \& Tremaine, S. 1987, {\sl Galactic Dynamics}, (Princeton University Press: Princeton)

\bibitem[Christlein \& Zaritsky(2008)]{christlein}
Christlein, D. \& Zaritsky, D. 2008, \apj, 680, 1053


\bibitem[Clowe et al.(2006)]{clowe}
Clowe, D., {Brada{\v c}}, M., {Gonzalez}, A.~H.,	{Markevitch}, M., {Randall}, S.~W., {Jones}, C. and {Zaritsky}, D.\ 2006, \apjl, 648, 109

\bibitem[Famaey \& Binney(2005)]{fb} 
Famaey, B., \&  Binney, J.\ 2005, \mnras, 363, 603

\bibitem[Fathi et al.(2007)]{fathi}
Fathi, K. et al. 2007, A\&A, 466, 905

\bibitem[Ferguson, et al.(1998)]{ferguson}
Ferguson, A.M.N., Wyse, R.F.G., Gallagher, J.S., and Hunter, D. A. 1998, \apj, 506, 19

\bibitem[Herbert-Fort et al.(2010)]{herbert-fort}
Herbert-Fort, S., Zaritsky, D., Christlein, D., \& Kannappan, S.J. 2010, \apj, in press (arXiv:1003.4536)

\bibitem[Hoekstra, Yee, \& Gladders(2004)]{hoekstra}
Hoekstra, H., Yee, H.K.C., \& Gladders, M.D. 2004, \apj, 606, 67

\bibitem[Kamphuis \& Briggs(1992)]{kamphuis}
Kamphuis, J., \& Briggs, F. 1992, A\&A, 253, 335

\bibitem[Mandelbaum et al.(2006)]{mandelbaum}
Mandelbaum, R., Hirata, C.M., Broderick, T., Seljak, U. \& Brinkmann, J. 2006, \mnras, 370, 1008

\bibitem[McGaugh \& de Blok(1998)]{mcgaugh}
McGaugh, S. \& de Blok, W.J.G. 1998, \apj, 499, 66

\bibitem[Merrifield(1992)]{merrifield}
Merrifield, M.R. 1992, \aj, 103, 1552

\bibitem[Milgrom(1983)]{milgrom}
Milgrom, M. 1983, \apj, 270 371

\bibitem[Natail et al.(1991)]{natali}
Natali, G., Pedichini, F., \& Righini, M. 1992, A\&A, 256, 79

\bibitem[Nipoti et al.(2007)]{2007MNRAS.379..597N} Nipoti, C.,
Londrillo, P., Zhao, H., \& Ciotti, L.\ 2007, \mnras, 379, 597

\bibitem[O'Brien, Freeman, \& van der Kruit(2010)]{obrien}
O'Brien, J. C., Freeman, K.C., and van der Kruit, P.C. 2010, A\&A, in press (arXiv:1003.3108)

\bibitem[Olling(1996)]{olling}
Olling, R.P. 1996, \aj, 112, 457

\bibitem[Rogstad \& Shostak(1972)]{rogstad}
Rogstad, D.H., \& Shostak, G.S. 1972, \apj 176, 315

\bibitem[Rubin, Ford, \& Thonnard(1978)]{rubin}
Rubin, V.C., Ford, W.K., Jr., \& Thonnard, N. 1978, \apjl, 255, 107

\bibitem[S{\'a}nchez-Salcedo et al.(2008)]{sanchez}
  S{\'a}nchez-Salcedo, F.~J., Saha, K., \& Narayan, C.~A.\ 2008,
  \mnras, 385, 1585
  
\bibitem[Sancisi \& Allen(1979)]{sancisi}
Sancisi, R., \& Allen, R.J. 1979, A\&A, 74, 73
  
\bibitem[Smith(1936)]{smith}
Smith, S. 1936, \apj, 83, 23

\bibitem[Thilker et al.(2005)]{thilker}
Thilker, D.A., et al. 2005, \apjl, 619, 79

\bibitem[Zaritsky \& Christlein(2007)]{zaritsky}
Zaritsky, D., \& Christlein, D., \aj, 134, 135

\bibitem[Zwicky(1933)]{zwicky}
Zwicky, F. 1933, AcHPh, 6, 110

\end{thebibliography}
\end{document}